\documentclass[aps,preprint,prd,showpacs,nofootinbib]{revtex4}
\usepackage{comment}
\usepackage{graphicx}
\usepackage{dcolumn}
\usepackage{bm}
\usepackage{hyperref}
\usepackage{graphicx} 
\usepackage{subfigure}
\usepackage{bm}
\usepackage{amsmath,amssymb}
\usepackage{latexsym}
\usepackage{ulem}
\usepackage{datetime}
\usepackage{scrtime}
\usepackage[dvipsnames, svgnames, x11names]{xcolor}

\definecolor{darkgreen}{RGB}{50,150,0}
\definecolor{purple}{cmyk}{0.5,0.75,0,0}
\definecolor{darkpurple}{RGB}{128,0,128}
\definecolor{ultramarine}{rgb}{0.07, 0.04, 0.56}
\definecolor{cadmiumgreen}{rgb}{0.0, 0.42, 0.24}
\definecolor{indigo(dye)}{rgb}{0.0, 0.25, 0.42}

\hypersetup{colorlinks=true,linkcolor=blue,citecolor=blue}

\def\be{\begin{equation}}
\def\ee{\end{equation}}
\def\ba{\begin{eqnarray}}
\def\ea{\end{eqnarray}}

\def\d{\mathrm{d}}

\begin{document}
\title{Constraints on Genesis Cosmology from the Smeared Null Energy Condition}
\author{Dong-Hui Yu$^{1}$}
\author{Mian Zhu$^{2}$}
\email[Corresponding author:~]{zhumian@scu.edu.cn}
\author{Yong Cai$^{1}$}
\email[Corresponding author:~]{caiyong@zzu.edu.cn}
\affiliation{$^1$ Institute for Astrophysics, School of Physics, Zhengzhou University, Zhengzhou 450001, China}
\affiliation{$^2$ College of Physics, Sichuan University, Chengdu 610065, China}

\begin{abstract}

The violation of the null energy condition (NEC) is essential for constructing nonsingular cosmological scenarios, such as Genesis cosmology, which avoids the initial singularity by initiating cosmic evolution from an asymptotically Minkowski state. To address theoretical concerns regarding the accumulation of negative energy, the smeared null energy condition (SNEC) has been proposed as a quantum-motivated, semi-local bound on NEC violation. In this work, we examine the implications of the SNEC conjecture for Genesis models, typically constructed within generalized Galileon theories. Our results demonstrate that SNEC imposes nontrivial restrictions on the viability of Genesis models, highlighting the SNEC conjecture as a powerful tool for constraining nonsingular cosmological scenarios.

\end{abstract}

\maketitle
\tableofcontents

\section{Introduction}
\label{sec:intro}

The persistence of certain energy conditions is crucial to the establishment of Penrose and Hawking's singularity theorems \cite{Penrose:1964wq,Hawking:1967ju,Hawking:1970zqf} and the positive mass theorem \cite{Schoen:1979zz}. Among various energy conditions, the null energy condition (NEC) and its violation are of particular importance in cosmology, see e.g. \cite{Rubakov:2014jja,Curiel:2014zba,Kontou:2020bta} for reviews. From a theoretical viewpoint, the Penrose singularity theorem, under the assumption of the NEC, establishes the null geodesic incompleteness of spacetime in gravitational collapse scenarios involving a trapped surface and appropriate global causal/compactness assumptions \cite{Penrose:1964wq}. Furthermore, without assuming any pointwise energy conditions, the Borde–Guth–Vilenkin argument shows that a spacetime with a positive past-averaged expansion rate is necessarily past geodesically incomplete \cite{Borde:1993xh, Borde:2001nh}. This past incompleteness poses certain challenges for specifying a physically reasonable initial state for primordial perturbations in inflationary cosmology \cite{Martin:2000xs,Cai:2019hge}.

Nonsingular cosmic bounce \cite{Gasperini:1992em,Finelli:2001sr,Piao:2003zm,Cai:2007qw,Easson:2011zy,Qiu:2011cy,Cai:2012va,Liu:2013kea,Koehn:2013upa,Qiu:2013eoa,Nojiri:2016ygo,Li:2016awk,Cai:2017pga,Ye:2019sth,Battista:2020lqv,Ilyas:2020qja,Battista:2022hqn,Zhu:2021whu,Zhu:2023lhv} and Genesis \cite{Creminelli:2010ba,Liu:2011ns,Wang:2012bq,Liu:2012ww,Creminelli:2012my,Hinterbichler:2012fr,Hinterbichler:2012yn,Easson:2013bda,Liu:2014tda,Pirtskhalava:2014esa,Nishi:2015pta,Cai:2016gjd,Nishi:2016ljg,Mironov:2019qjt,Ilyas:2020zcb,Zhu:2021ggm,Cai:2022ori} are two representative scenarios for addressing the past geodesic incompleteness of inflation, and both involve violating the NEC. Theoretical studies in general relativity indicate that the geometric structure of a traversable wormhole requires matter fields that violate the NEC to sustain the throat \cite{Morris:1988cz,Morris:1988tu}.
Recently, observations of baryon acoustic oscillations by the DESI collaboration \cite{DESI:2025fii} have also sparked interest in whether dark energy in the late-time universe violates the NEC \cite{Ye:2025ulq,Moghtaderi:2025cns,Caldwell:2025inn}.
In addition, NEC violation also provides a compelling mechanism for the formation of PBHs as a dark matter candidate \cite{Cai:2023uhc}, as well as for generating stochastic gravitational-wave background signals detectable by pulsar timing array observations \cite{Zhu:2023lbf,Ye:2023tpz}, see also \cite{Cai:2020qpu,Cai:2022nqv,Cai:2022lec,Pan:2024ydt,Li:2024oru,Li:2025ilc,Lai:2025xov,Lai:2025efh}.

Although widely employed in theoretical models, the viability of NEC violation in the real world remains in question. Under the perfect-fluid assumption, NEC violation means the total energy density plus pressure becomes negative along a null direction, which allows exotic behaviors like an increase of energy density even in an adiabatically expanding universe. For theorists, NEC violation is often viewed unfavorably because it tends to trigger pathological instabilities, including ghost and gradient instabilities \cite{Libanov:2016kfc,Kobayashi:2016xpl}, see also \cite{Battarra:2014tga,Koehn:2015vvy,Ijjas:2016tpn}. 
It is demonstrated in the context of nonsingular cosmology that fully stable
NEC violation can be constructed in ``beyond Horndeski'' theories \cite{Cai:2016thi,Creminelli:2016zwa,Cai:2017tku,Cai:2017dyi,Kolevatov:2017voe}.
Furthermore, the NEC can notoriously be violated at the quantum level, for instance in the quantum Casimir effect \cite{Brown:1969na} and in the Hawking evaporation process (see, e.g. \cite{Kontou:2023ntd} for a recent update on this topic). Notably, in the cosmological context, NEC violation can also emerge from the secular growth of vacuum expectation values for self-interacting scalar fields on a de Sitter background, as derived within the framework of perturbative quantum field theory \cite{Kahya:2009sz}.

Nevertheless, despite that the NEC may not be strictly forbidden, the physical necessity of bounding the accumulation of negative energy along null geodesics motivates the study of the averaged null energy condition (ANEC) \cite{Borde:1987qr,Klinkhammer:1991ki, Ford:1994bj,Fewster:2006uf,Graham:2007va} and the smeared null energy condition (SNEC) \cite{Freivogel:2018gxj}. The ANEC conjectures that the quantum expectation value of the null energy must be non-negative when averaged over an entire null (achronal) geodesic. 
Although the ANEC is well-motivated from the perspective of quantum gravity, its definition makes it a global and highly non-local concept, which limits its applicability in cosmology. Because cosmological NEC violation typically occur only over a finite, specific epoch, the ANEC fails to capture the local or semi-local nature of NEC violation in cosmology.

Inspired by the limitation of the ANEC, SNEC is proposed as a quantum-motivated semi-local energy condition, conjectured to hold at the level of semi-classical gravity \cite{Freivogel:2018gxj}. The idea is to replace the integral average in ANEC by a weighted average through a smearing function, similar to a window function that regulates NEC in a particular regime. With the localization provided by the smearing function, the SNEC can be applied to generic features of NEC violation in gravitational theories \cite{Freivogel:2020hiz,Fliss:2023rzi,Fliss:2024dxe} as well as to NEC violation in cosmology \cite{Moghtaderi:2025cns}. In particular, Ref. \cite{Moghtaderi:2025cns} studies the NEC violation in the context of dark energy and nonsingular bouncing cosmology. By assuming that SNEC hold fundamentally, they arrive at upper bounds on model parameters of those NEC-violating cosmological scenarios. This shows that the SNEC conjecture can indeed serve as a powerful tool for studying NEC-violating cosmological scenarios.

In this paper, we apply SNEC conjecture to another simple-yet-significant non-singular cosmology scenario, the Genesis cosmology \cite{Creminelli:2010ba}. The idea traces back to Ref. \cite{Piao:2003ty}, where it is proposed that the observed nearly scale-invariant curvature power spectrum \cite{Planck:2018vyg, Planck:2018jri} can originate from a slowly expanding phase of the universe. Genesis cosmology evades the initial cosmological singularity by initiating cosmic evolution from an asymptotically Minkowski state and undergoes a stable phase of NEC violation that allows the scale factor to start expanding without ever reaching a singularity. The Genesis model is concretely constructed in the context of Generalized Galileon theory \cite{Deffayet:2011gz,Nishi:2015pta}, a subclass of Horndeski theory \cite{Horndeski:1974wa}. The robustness of the NEC-violating Genesis scenario remains an active field to investigate in recent years \cite{Ageeva:2020gti,Ageeva:2020buc,Ilyas:2020zcb,Zhu:2021ggm,Cai:2022ori}. We are motivated to examine whether the SNEC conjecture can impose any constraints on the Genesis scenario.

This paper is organized as follows. We briefly introduce the SNEC conjecture and the Genesis cosmology in Secs. \ref{sec:SNEC} and \ref{sec:Genesis}, respectively. Our main result, the constraints on Genesis cosmology from SNEC, is presented in Sec. \ref{sec:constraint}. We summarize our results in Sec. \ref{sec:conclusion}.
Throughout this paper, we work in Planck units such that $c = 1$ and $M_{\text{P}}\equiv (8\pi G )^{-1/2}$. We adopt the mostly positive metric signature $(-,+,+,+)$.

\section{The SNEC conjecture}
\label{sec:SNEC}

\subsection{The NEC}

Classically, the NEC is expressed as 
\begin{equation}
\label{eq:NEC}
    T_{\mu \nu} k^{\mu} k^{\nu} \geq 0 ~,
\end{equation}
where $T_{\mu \nu}$ is the stress-energy tensor, $k^\mu$ is an null vector satisfying $k_\mu k^\mu = 0$. The violation of NEC indicates that $T_{\mu \nu} k^{\mu} k^{\nu} < 0$. The NEC states that energy moving in null directions is non-negative. Provided that the Einstein equatioin holds, i.e., $G_{\mu\nu}\equiv  R_{\mu \nu} -g_{\mu\nu}R/2=M_{\text{P}}^{-2} T_{\mu \nu}$, the NEC is equivalent to the null convergence condition (NCC)
\be \label{eq:NCC} 
R_{\mu\nu}k^\mu k^\nu\geq 0\,, 
\ee 
which can be regarded as the geometric form of the NEC, see e.g. \cite{Rubakov:2014jja,Curiel:2014zba,Kontou:2020bta} for reviews. 

When combined with the Raychaudhuri equation, the NCC implies that the expansion rate of a null congruence, $\theta$, must be non-increasing along the geodesic, i.e., $\frac{d\theta}{d\lambda} \leq 0$. Intuitively, this result indicates that the NEC requires gravity to be a purely attractive force for light rays, causing null geodesics to focus rather than diverge. This property is crucial for establishing the singularity theorem. Notably, when gravity is modified, the NEC (\ref{eq:NEC}) is not necessarily equivalent to the NCC (\ref{eq:NCC}). In the literature, it is often customary to use the term ``NEC'' to refer either to the NCC itself or to a modified form of the NEC that becomes equivalent to the NCC in the given modified gravity theory, although this is not always the case.

\subsection{The SNEC conjecture}

The SNEC introduces a smearing function $f_{\sigma}(\lambda)$ for a null geodesic $\gamma$,  with $\d \lambda=a \d t$, where $\lambda$ is the affine parameter of $\gamma$, and $\sigma$ denotes the smearing scale. 
The smearing function serves as a window function. It must be positive, square integrable, continuously differentiable, and normalized:
\begin{equation}
\label{unity}
    \int_{\gamma} \d\lambda f_{\sigma}^2(\lambda) = 1 ~.
\end{equation}
In this sense, we can define a functional 
\begin{equation} \label{eq:251204-1}
    \mathbb{E}_{\sigma}(F) \equiv \int_{\gamma} \d\lambda f_{\sigma}^2(\lambda)  F(\lambda) ~,
\end{equation}
to evaluate the smearing of a test function $F(\lambda)$ along the geodesics. 

When the smearing function is locally distributed, for example, when the smearing function takes a Gaussian form 
\begin{equation}
\label{eq:smearGaussian}
    f^2_{\sigma} (\lambda) = \frac{1}{\sqrt{2\pi} \sigma} \exp \left( -\frac{( \lambda - \bar{\lambda} )^2}{2\sigma^2} \right) ~,
\end{equation}
or a a Lorentzian form
\be \label{eq:smearLorentzian}
f_{\sigma}^2(\lambda)=\frac{\sigma}{\pi\sqrt{2}((\lambda-\bar{\lambda})^2+\sigma^2/2)}~,
\ee
the functional $\mathbb{E}_{\sigma}(F)$ extracts the local feature of $F(\lambda)$ around $\lambda = \bar{\lambda}$ with a characteristic smearing scale $\sigma$, which is conventionally defined by
\begin{equation}
\label{eq:scale}
    \frac{1}{\sigma^2} \equiv 4 \int_{\gamma} \d\lambda \left( \frac{\d f_{\sigma}}{\d\lambda} \right)^2 ~.
\end{equation}
This definition is self-consistent for both the Gaussian and Lorentzian forms above.



With the smearing process, one may expect the semi-classical version of SNEC being
$\mathbb{E}_{\sigma} \left[ \langle \Psi | \hat{T}_{\mu \nu} k^{\mu} k^{\nu} | \Psi \rangle \right] \geq 0$ for any quantum state $| \Psi \rangle$.
However, a counter example exists even in the standard quantum field theory \cite{Krommydas:2017ydo}. Then, a new bound in semi-classical version, known as the SNEC, is proposed in Ref. \cite{Freivogel:2018gxj}. The SNEC conjectures that for all states $| \Psi \rangle$ within the regime of perturbative quantum gravity, in any dimension $d$ and for locally flat regions, the expectation value of the null-null contracted stress-energy tensor, when smeared over any achronal null geodesics, is bounded by
\begin{equation}
\label{eq:SNEC}
 \mathbb{E}_{\sigma} \left[ \langle \Psi | \hat{T}_{\mu \nu} k^{\mu} k^{\nu} | \Psi \rangle \right] \geq - \frac{8\pi M_{\text{P}}^2}{\sigma^2} B ~, 
\end{equation}
where $B$ is a dimensionless constant. A benchmark value of $B$ is $B=1/(32\pi)$ \cite{Leichenauer:2018tnq}.
However, we should keep in mind that $B$ 
is yet to be determined in principle. 

In practice, for scenarios of the primordial universe, it is often more convenient to rewrite the condition \eqref{eq:SNEC} in the following geometric form
\begin{equation}
    \label{eq:SNECR}
\mathbb{E}_{\sigma} \left[ \langle \Psi | R_{\mu \nu} k^{\mu} k^{\nu} | \Psi \rangle \right] \geq - \frac{8\pi B}{\sigma^2} \,,
\end{equation}
as long as the semi-classical Einstein equation $G_{\mu\nu}=M_{\text{P}}^{-2}\langle \Psi | \hat{T}_{\mu \nu} | \Psi \rangle$ remains valid. For a spatially-flat FLRW universe, the line element is $\d s^2 = -\d t^2 + a(t)^2 \d \vec{x}^2$, which gives $R_{\mu \nu} k^{\mu} k^{\nu} = - 2{\dot{H}}/{a^2}$.
According to Eq. (\ref{eq:251204-1}), the SNEC conjecture (\ref{eq:SNECR}) can be rewritten as 
\begin{equation}
\label{eq:SNECdt}   \int_{t_\mathrm{s}}^{t_{\mathrm{e}}}\mathrm{d} t f_{\sigma}^2(\lambda(t))\frac{\dot{H}}{a}\le \frac{4 \pi B}{\sigma^2}\,,
\end{equation}
where $t_\mathrm{s}$ and $t_\mathrm{e}$ denote the beginning and the end of the cosmic interval we are interested in, respectively. Here, we have already used $\d\lambda=a \d t$. 
It is interesting to note that the secular growth of $\langle T_{\mu\nu}k^\mu k^\nu\rangle \approx -\frac{4}{3}\frac{\lambda H^4}{(4\pi)^4}\ln a(t)$ for a massless, minimally coupled scalar with quartic self-interaction on a non-dynamical de Sitter background (as computed in \cite{Kahya:2009sz} with appropriate initial state corrections) respects the SNEC, even though it violates the classical NEC (see the Appendix. \ref{Sec:app1} for details).
In the following, (\ref{eq:SNECdt}) serves as the master condition for examining the Genesis scenario.



\section{Genesis cosmology}
\label{sec:Genesis}

In this section, we briefly introduce the Genesis scenario and clarify the specific models we consider.

\subsection{A brief review of cosmological perturbation theory}
\label{sec:CPT}


In cosmology, the most important observational result is the nearly scale-invariant power spectrum of curvature perturbation $\zeta$ on large scales:
\begin{equation}
    \mathcal{P}_{\zeta}(k) = A_s \left( \frac{k}{k_{\ast}} \right)^{n_s - 1} ~,
\end{equation}
where $A_s = 2.1 \times 10^{-9}$ from Planck collaboration \cite{Planck:2018jri} and $n_s = 0.9709 \pm 0.0038$ from the recent Atacama Cosmology Telescope results \cite{ACT:2025fju}. 

To evaluate the curvature power spectrum
\begin{equation}
\label{eq:Pzetadef}
     \langle \zeta_{\vec{k}} \zeta_{\vec{p}} \rangle = (2\pi)^3 \delta(\vec{k} + \vec{p}) \frac{2\pi^2}{k^3} \mathcal{P}_{\zeta}(k) ~,
\end{equation}
one needs to know the evolution of $\zeta$ from the quadratic action
\begin{equation}
    S_{\zeta}^{(2)} = \int \d\tau \d^3x \frac{z_s^2}{2} \left[ \zeta^{\prime 2} - c_s^2 (\partial_i \zeta)^2 \right] ~,
\end{equation}
which leads to the well-known Mukhanov-Sasaki (MS) equation
\begin{equation}
\label{eq:MS}
    v_k^{\prime \prime} + \left( c_s^2 k^2 - \frac{z_s^{\prime \prime}}{z_s} \right) v_k = 0 ~.
\end{equation}
In the following, we use a prime to denote the differentiation with respect to the conformal time $\tau$, $\d\tau \equiv \d t/a$. 

When matter field is minimally coupled to gravity, one has $z_s \propto a$ and accordingly
\begin{equation}
    \frac{z_s^{\prime \prime}}{z_s} = \frac{a^{\prime \prime}}{a} = 2a^2 H^2 + a H^{\prime}  = \mathcal{H}^{\prime} + \mathcal{H}^{2} ~,
\end{equation}
where we defined the Hubble parameter $H \equiv \frac{1}{a} \frac{da}{dt}$ and the conformal Hubble parameter $\mathcal{H} \equiv a^{\prime}/a = aH$. In addition, it is generically assumed that the scale factor is a polynomial of conformal time $\tau$, i.e.,
\begin{equation}
\label{eq:ageneral}
    a = a(\tau_{\ast}) \left( \frac{\tau}{\tau_{\ast}} \right)^n ~\to~ \mathcal{H} = \frac{n}{\tau} ~,
\end{equation}
where $\tau_{\ast}$ is a scaling time. The parameterization turns the MS equation into
\begin{equation}
    v_k^{\prime \prime} + \left( c_s^2 k^2 - \frac{n(n-1)}{\tau^2} \right) v_k = 0 ~,
\end{equation}
whose general solution is the Hankel function 
\begin{equation}
    v_k(\tau) = c_1 H_{\nu}^{(1)} (c_s k\tau) + c_2 H_{\nu}^{(2)} (c_s k\tau) ~,~ \nu = n - \frac{1}{2} ~.
\end{equation}
Imposing the vacuum initial condition which picks up the $H_{\nu}^{(1)}$ branch of solution, one may see that the scale-invariant power spectrum can be acquired when $\nu = \pm 3/2$. Namely, $n = -1$ in an expanding universe, or $n = 2$ in a contracting universe \cite{Wands:1998yp}. In the former case, the universe exponentially expands in cosmic time, corresponding to the well-known inflation scenario; in the latter case, the universe contracts as if a pressure-less dust dominates the whole matter content, known as the matter bounce scenario. 

\subsection{Scale-invariance from a slowly expansion state}
\label{sec:SI}

In Ref. \cite{Piao:2003ty}, an alternative ansatz for cosmic evolution is considered, i.e.,
\begin{equation}
\label{eq:aGenesis}
    a(\tau) = 1 + \frac{H_0}{2\alpha} \frac{(-\tau_0)^{2\alpha + 1}}{(-\tau)^{2\alpha}} ~,~ -\infty < \tau < \tau_0 < 0 ~,
\end{equation}
where we adopted the convention in Ref. \cite{Nishi:2015pta}. This ansatz describes a slowly expanding phase, starting from a quasi-Minkowskian configuration, which is known as the Genesis phase. As long as $a$ does not change significantly, the Hubble parameter becomes
\begin{equation}
\label{eq:HGenesis}
    \mathcal{H} \simeq H_0 \left( \frac{-\tau_0}{-\tau} \right)^{2\alpha + 1} ~,
\end{equation}
which vanishes when $\tau \ll \tau_0$. Apparently, the quasi-Minkovskian configuration cannot eternally last, as the ansatz \eqref{eq:aGenesis} blows up at $\tau = 0$. We set $\tau = \tau_0$ as the termination of quasi-Minkovskian configuration, thus $H_0$ labels the Hubble parameter at the end of Genesis phase. Notably, in the Genesis phase, the scale factor $a$ is nearly constant. This allows us to set the cosmic time $t$ approximately equal to the conformal time $\tau$ (i.e., $t \simeq \tau$), which in turn implies $H \simeq \mathcal{H}$. For convenience, we will not distinguish between quantities defined in cosmic time and those defined in conformal time throughout the subsequent analysis.

In a flat FLRW universe, $R_{00} = -3 \mathcal{H}^{\prime}$, $R_{ij} = (2 \mathcal{H}^{2} + \mathcal{H}^{\prime}) \delta_{ij}$.
Taking an outgoing null geodesic $k^{\mu} (\tau,r,\theta,\varphi) = a^{-2} (1,1,0,0)$ in a spherically coordinate, it is easy to obtain
\begin{equation}
\label{eq:NCCRKK}
    R_{\mu \nu} k^{\mu} k^{\nu} = - 2\frac{\dot{H}}{a^2} \simeq - 2\frac{\mathcal{H}^{\prime}}{a^3} \simeq -2 \mathcal{H}^{\prime} = -2H_0(2\alpha + 1) \frac{(-\tau_0)^{2\alpha + 1}}{(-\tau)^{2\alpha + 2}} < 0 ~.
\end{equation}
Namely, the NEC is violated in the Genesis phase. 

Utilizing the expression of $H$, the MS equation approximates to
\begin{equation}
    v_k^{\prime \prime} + \left[ c_s^2 k^2 - (2\alpha + 1) H_0 \frac{(-\tau_0)^{2\alpha + 1}}{(-\tau)^{2\alpha + 2}} \right] v_k = 0 ~\to~ n_s = 5 - 2\alpha 
\end{equation}
when $\alpha > 1/2$, or
\begin{equation}
    v_k^{\prime \prime} + \left[ c_s^2 k^2 - H_0^2 \left( \frac{-\tau_0}{-\tau} \right)^{4\alpha + 2} \right] v_k = 0 ~,~ n_s = 2\alpha + 3 
\end{equation}
when $0 < \alpha < 1/2$. It is easy to see that a scale-invariant curvature power spectrum is acquired when $\alpha = 2$.

When $\alpha \neq 2$, the scale-invariant power spectrum can be acquired via the curvaton mechanism \cite{Lyth:2001nq}. For instance, in the original Galileon Genesis model \cite{Creminelli:2010ba}, the authors set $\alpha = 1$ and argue that the interaction with Galileon field naturally leads to the scale-invariance of curvaton field, see also Ref. \cite{Wang:2012bq} on an update on this topic. Additionally, in Ref. \cite{Cai:2016gjd}, a slow expansion model for the case $\alpha = 4$ was proposed by introducing a non-minimal coupling. This model results in a nearly scale-invariant power spectrum for primordial gravitational waves (GWs). However, the corresponding primordial curvature perturbation exhibits a blue tilt, necessitating the introduction of an auxiliary light field to achieve nearly scale-invariant scalar perturbations. In light of this fact, we will focus on two representative models of Galileon Genesis, the $\alpha = 2$ model where the scale-invariance comes from vacuum fluctuations of Galileon fields, and the $\alpha = 1$ model, which serves as a representative of Genesis scenarios employing the curvaton mechanism.

\subsection{The model of Galileon Genesis}
As proposed in Ref. \cite{Creminelli:2010ba} and generalized in Ref. \cite{Nishi:2015pta}, the Genesis phase can be concretely realized with a Galileon field $\phi$. Here we take the action to be
\begin{equation}\label{eq:LGenesis}
    S= \int \d^4x \sqrt{-g} \left[ M_{\text{P}}^2\frac{R}{2} + f_1(\phi) X + f_2(\phi) \frac{X^2}{M^4} - G(\alpha) \frac{X^{\alpha}}{M^{4\alpha - 1}} \Box \phi \right]~,
\end{equation}
where $X \equiv -\nabla_{\mu} \phi \nabla^{\mu} \phi/2$ is the canonical kinetic term of $\phi$, and $\Box \equiv \nabla_{\mu} \nabla^{\mu}$. We set the field $\phi$ to have the dimension of mass, $[\phi] = [M]$. Consequently, the kinetic term $X$ carries the dimension of $ [M]^2$, and $\Box \phi$ has the dimension of $ [M]^3$. To ensure dimensional consistency, we introduce a mass scale $M$, $M<M_{\text{P}}$ \cite{Cai:2022ori}. As a result, $f_1(\phi)$, $f_2(\phi)$ and $G(\alpha)$ are dimensionless functions. 

The Friedmann equations can be obtained as
\begin{equation}
    3H^2 M_{\text{P}}^2 = Xf_1+ 3\frac{X^2}{M^4} f_2 + 6\alpha G(\alpha) H \dot{\phi} \frac{X^{\alpha}}{M^{4\alpha-1}} ~,
\end{equation}
\begin{equation}
    \dot{H} M_{\text{P}}^2 = \alpha G(\alpha) \frac{X^{\alpha}}{M^{4\alpha - 1}} \left( \ddot{\phi} - 3H \dot{\phi} \right) - Xf_1 M_{\text{P}}^2 - 2\frac{X^2}{M^4} f_2 ~.
\end{equation}
When the auxiliary function $f_1$ and $f_2$ has the following form
\begin{equation}
      f_1 \propto e^{2\alpha \phi/M} ~,~ f_2 \propto e^{2(\alpha - 1) \phi/M} ~,
\end{equation}
the action admits the following attractor solution
\begin{equation}
e^{\lambda \phi/M} \simeq \frac{{\rm const.}}{-t} ~,~ H \simeq \frac{{\rm const.}}{(-t)^{2\alpha + 1}} ~. 
\end{equation}
This fact enables us to take the following parameterizations for the model constructions of the cases $\alpha = 1$ and $\alpha = 2$. 

\subsubsection{Case I: $\alpha = 1$}

In the case of $\alpha = 1$, we set these functions in the action (\ref{eq:LGenesis})
\begin{equation}
f_1(\phi)=-2\lambda_1e^{2\phi/M}~,~f_2(\phi)=4\lambda_2~,~ G(\alpha) = \lambda_g ~.
\end{equation}
As a result, the Friedmann equations become
\begin{equation}
    3M_{\text{P}}^2H^2=-\lambda_1e^{2\phi/M}\dot{\phi}^2+\frac{3\lambda_2}{M^4}\dot{\phi}^4+\frac{3\lambda_g}{M^3}H\dot{\phi}^3~\,,
\end{equation}
\begin{equation}   \dot{H}M_{\text{P}}^2=\lambda_1e^{2\phi/M}\dot{\phi}^2-\frac{2\lambda_2}{M^4}\dot{\phi}^4-\frac{3\lambda_g}{2M^3}H\dot{\phi}^3+\frac{\lambda_g}{2M^3}\dot{\phi}^2\ddot{\phi}~.
\end{equation}
For the Genesis solution, we have
\begin{equation}
    e^{2\phi/M}=\frac{3\lambda_2}{\lambda_1}\frac{1}{M^2(-t)^2}\to\dot{\phi}=\frac{M}{-t}\,.
\end{equation}
The Friedmann equations indicates $H\simeq\frac{2\lambda_2+\lambda_g}{6M_{\text{P}}^2(-t)^3}$ and
\begin{equation}
\label{eq:Halpha1}
    \dot{H}\simeq\frac{2\lambda_2+\lambda_g}{2M_{\text{P}}^2(-t)^4} \,.
\end{equation}

\subsubsection{Case II: $\alpha = 2$}
In the case of $\alpha = 2$, we set
\begin{equation}
    f_1(\phi) = - \frac{3}{2} \kappa^2 e^{4\phi/M} ~,~ f_2(\phi) = e^{2\phi/M} ~,~ G(\alpha) = \gamma ~.
\end{equation}
The Friedmann equations are simplified to 
\begin{equation}
    3M_{\text{P}}^2H^2 = - \frac{3}{4} \kappa^2 e^{4\phi/M} \dot{\phi}^2 + \frac{3}{4} \frac{\dot{\phi}^4}{M^4} e^{2\phi/M} + 3 \gamma H \frac{\dot{\phi}^5}{M^7}  ~,
\end{equation}
\begin{equation}
    \dot{H}M_{\text{P}}^2 = \frac{\gamma}{2} \frac{\dot{\phi}^4}{M^7} \left( \ddot{\phi} - 3H \dot{\phi} \right) + \frac{3}{4} \kappa^2 e^{4\phi/M} \dot{\phi}^2 - \frac{1}{2} 
    \frac{\dot{\phi}^4}{M^4} e^{2\phi/M} ~.
\end{equation}
The Genesis solution indicates that
\begin{equation}
    e^{\phi/M} = \frac{1}{\kappa (-t)M} ~\to~ \dot{\phi} = \frac{M}{-t} ~.
\end{equation}
Using the Friedmann equations, we get
\begin{equation}
\label{eq:Halpha2}
   \dot{H} \simeq \frac{1+2\kappa^2 \gamma}{4\kappa^2M_{\text{P}}^2M^2} \frac{1}{(-t)^6} ~.
\end{equation}
It is easy to see that Eq. \eqref{eq:Halpha2} meets our parameterization \eqref{eq:HGenesis}. Aadditionally, the sound speed squred of scalar perturbation reads $c_s^2 = (8\kappa^2 \gamma - 1)/3$. In order to avoid the gradient instability, we should require that
\begin{equation}
    8\kappa^2 \gamma > 1 ~.
\end{equation}



\section{Constraints on Genesis cosmology from SNEC}
\label{sec:constraint}

In this section, we discuss the constraints on Genesis cosmology from SNEC conjecture. As discussed in Sec. \ref{sec:SI}, we will consider two models of Galileon Genesis, with $\alpha = 1$ and $\alpha = 2$ respectively.
Given that we have obtained the background solutions for $H(t)$ and $\dot{H}(t)$ for both cases, we can now conveniently use the SNEC bound, expressed by Eq. (\ref{eq:SNECdt}), to constrain the parameter space of the Genesis model.

We should note the choice of the integration limits, $t_\mathrm{s}$ and $t_\mathrm{e}$, in Eq. (\ref{eq:SNECdt}). 
In the Genesis scenario, the universe undergoes an NEC-violating Genesis phase, which is followed by reheating and a transition into a radiation-dominated era. Since our goal is to constrain the model parameters using the SNEC, we focus primarily on the NEC-violating stage. In this context, the null geodesics are past-complete ($t_{\mathrm{s}}=-\infty$) but future-incomplete. 
Therefore, we can set the initial time $t_\mathrm{s}$ to the infinite past ($t_\mathrm{s} \to -\infty$). The final time $t_\mathrm{e}$ should be taken as the time when the Genesis phase ends, which will be specified in detail below.
To simplify the expressions, we adopt the convention $M_{\text{P}}=1$ for the remainder of this paper. 



\subsection{Case I: $\alpha = 1$}

We can obtain the scale factor through \eqref{eq:Halpha1}, which results in
\begin{equation}
\label{eq:aalpha1}
    a(t)=e^{\int H \mathrm{d} t}=\mathrm{exp}\left[\frac{2\lambda_2+\lambda_g}{12}\frac{1}{(-t)^2} \right]\simeq1+\frac{2\lambda_2+\lambda_g}{12}\frac{1}{(-t)^2}~\,,
\end{equation}
for $|t|\gg1/M$ \cite{Cai:2022ori}. Since the Genesis background asymptotically approaches Minkowski space in the infinite past, we have already set $a(-\infty)=1$ in the above expression.
In the following calculations, we will use a series expansion of $a(t)$ given by Eq. (\ref{eq:aalpha1}), which introduces only a small error under the parameter choices considered below.


In order to utilize Eq. (\ref{eq:SNECdt}), we need to specify a physical end time beyond which the approximation $|t|\gg1/M$ is no longer valid. We denote this cutoff time as $t_{\mathrm{e}}$ and set it to
    \begin{equation}
        t_{\mathrm{e}}=-\frac{1}{M}.
    \end{equation}
To ensure proper normalization, the smearing function must satisfy the condition given by Eq. (\ref{unity}), i.e.,
    \begin{equation}
        \int_{-\infty}^{-1/M}\mathrm{d} t \, a(t)f^2_{\sigma}(\lambda(t))=1\,.
    \end{equation}
    
The parameter $\bar{\lambda}=\lambda(\bar{t})$ controls the central position of the smearing function. Similarly, the width of the smearing is written as $\sigma=[\lambda(t_+)-\lambda(t_-)]/2$, where $t_-=\bar{t}-\Delta t/2$. Due to the presence of the cutoff time $t_{\mathrm{e}}$, the upper limit $t_+$ must be defined as a piecewise function, i.e.,
\begin{equation}
    \label{eq:t_+}
    t_+= \left\{\begin{matrix}
    \bar{t}+\Delta t/2\,,  \quad \text{if}\quad \bar{t}+\Delta t/2\le t_{\mathrm{e}}\,, \\
    t_{\mathrm{e}}\,,\quad~~~~~~~~~ \text{if}\quad \bar{t}+\Delta t/2> t_{\mathrm{e}}.
    \end{matrix}\right.
\end{equation}
Under such parameter settings, $\bar{t}$ controls the position of the smearing, while $\Delta t$ determines the scale of the smearing function.

Prior to numerical implementation, we analyze the SNEC's mathematical structure and physical interpretation. Notably, the SNEC inherently bounds the magnitude of allowable NEC violations, as established earlier.
The magnitude of the NEC violation is directly reflected in $\dot{H}$. The violation of the NEC manifests unambiguously as $\dot{H} > 0$, i.e.,
    \begin{equation}
        \lambda_2+\frac{\lambda_g}{2}>0.
    \end{equation}

The Genesis solution given by Eq. \eqref{eq:Halpha2} reveals a symmetric dependence on $\lambda_2$ and $\frac{\lambda_g}{2}$. This symmetry implies that constraining the magnitude of NEC violation effectively imposes an upper bound on the linear combination $\lambda_2 + \frac{\lambda_g}{2}$. To constrain either parameter individually, one simply needs to fix the other as a constant. Thus , the combination $\lambda_2 + \frac{\lambda_g}{2}=\Lambda$ emerges as the fundamental parameter governing NEC violation in this framework, thereby providing a more compact parametrization and making the underlying physics more transparent. 

\begin{figure}[h]
        \centering
        \includegraphics[width=0.5\linewidth]{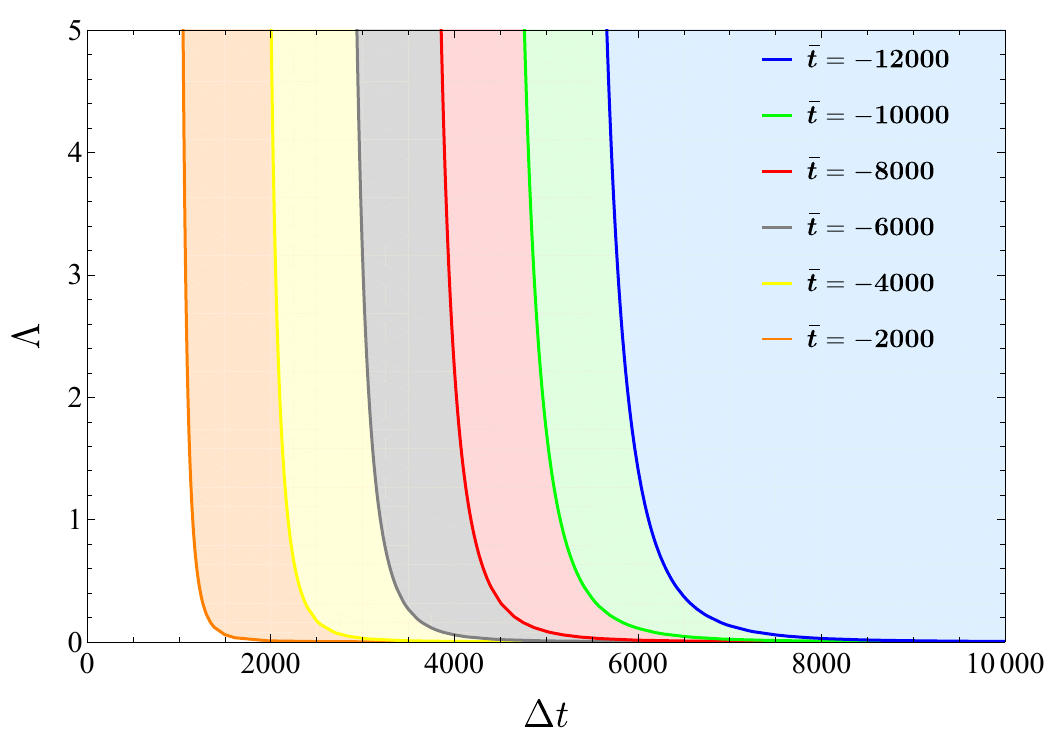}
        \caption{The SNEC constraints with $B=1/(32\pi)$ and $M=0.9$ fixed. The shaded regions represent the parameter space in which the SNEC is violated.}
        \label{F1}
\end{figure}

\begin{figure}[htbp]
    \subfigure[~~$\bar{t}=-10000,M=0.9$]{\includegraphics[width=.48\textwidth]{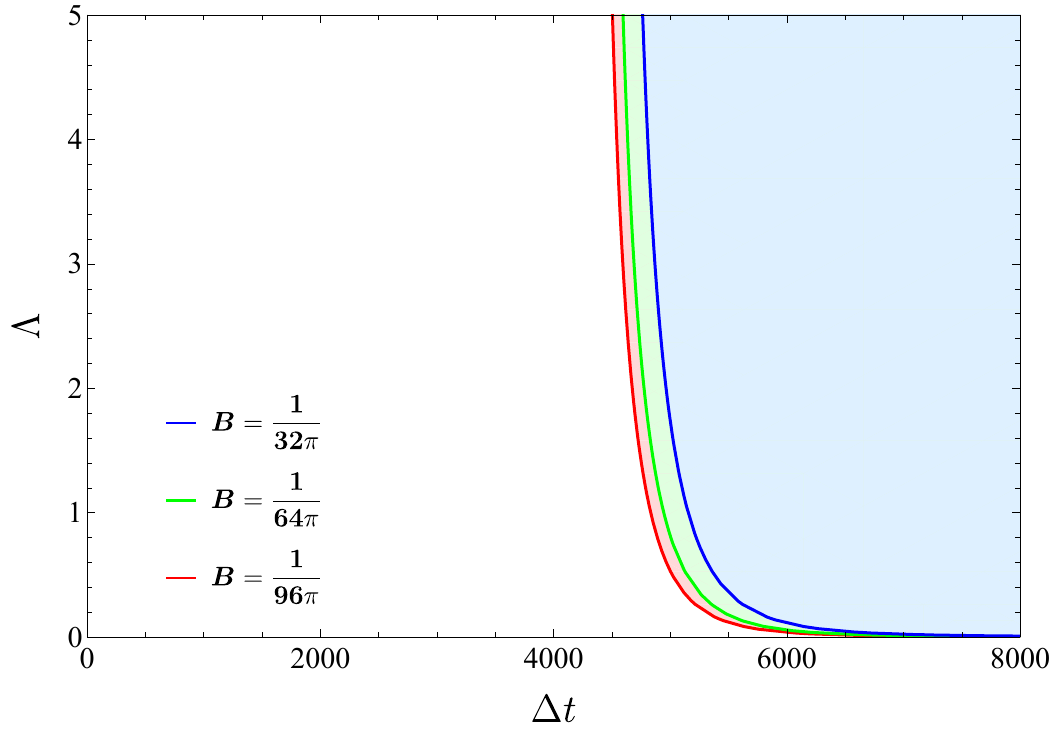} }
    \subfigure[~~$\bar{t}=-50000,B=1/(32\pi)$]{\includegraphics[width=.49\textwidth]{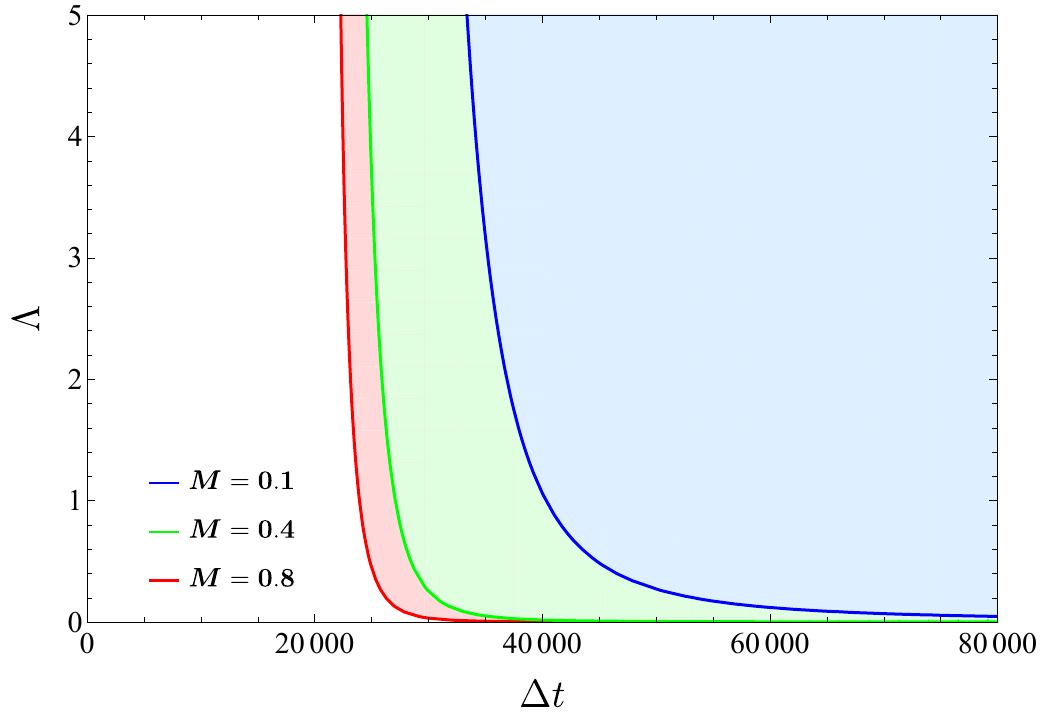} }
    \caption{The SNEC constraints with $\bar{t}$ and $M$ fixed (left channel), and with $\bar{t}$ and $B$ fixed (right channel). The shaded regions represent the parameter space in which the SNEC is violated.}\label{F2}
    \end{figure}

In this part, we adopt a Gaussian smearing function and numerically evaluate the SNEC expression (Eq. (\ref{eq:SNECdt})). The result depends on the values of $\bar{t}$, $\Delta t$, $B$, $\Lambda$, and $M$.
In Fig. \ref{F1}, we present the constraint on $\Lambda$ as a function of $\Delta t$, with different choices of $\bar{t}$, while the other parameters are fixed as $B=1/(32\pi)$ and $M=0.9$. A similar plot is shown in Fig. \ref{F2}, where $B$ and $M$ is varied.
We first observe that the constraint on $\Lambda$ increases monotonically with increasing $\Delta t$. This result is not surprising. The parameter $\Delta t$ controls the width of the smearing function, and as $\Delta t$ becomes larger, the smearing scale $\sigma$ also increases. This effectively reduces the allowed magnitude of NEC violation. Since the entire Genesis cosmology corresponds to a NEC-violating phase, a longer smearing scale naturally leads to tighter constraints.

We now discuss the impact of $\bar{t}$ on the SNEC constraint on $\Lambda$. It can be seen that, as $\bar{t}$ increases (i.e., its absolute value decreases), the SNEC places a tighter constraint on $\Lambda$. This indicates that the SNEC is more sensitive to the later stages of the Genesis evolution.

This behavior can be well understood. Recall that the measure of NEC violation, $\dot{H}$, increases monotonically with time in the Genesis scenario-as $t$ increases (in absolute value decreases), $\dot{H}$ becomes larger. This means the NEC violation becomes stronger as the universe evolves. Consequently, regions closer to the end of the Genesis phase contribute more significantly to the SNEC integral. Placing the peak of the smearing function in regions with stronger NEC violation naturally leads to tighter constraints on $\Lambda$.
The effect of $B$ on the SNEC constraint is also straightforward. As $B$ decreases, the constraint on $\Lambda$ becomes more stringent, which follows directly from the SNEC inequality. The significance of analyzing $B$ lies in the fact that it remains an undetermined constant. If observations can provide bounds on $\Lambda=\lambda_2+\frac{\lambda_g}{2}$, this may, in turn, place indirect constraints on the value of $B$.

As for the dependence on $M$, we find that smaller values of $M$ correspond to an earlier termination time $t_{\mathrm{e}}$. As a result, a portion of the highly NEC-violating region is truncated from the integral. As noted earlier, since $\dot{H}$ is increasing, this truncation reduces the contribution from regions where NEC violation is strongest, thereby weakening the constraint.

If we consider a single variable (such as $\lambda_2$
or $\lambda_g$) by fixing the other, it is evident that the result would not differ significantly from the case shown in the figures. The difference would merely amount to a constant shift.

To better illustrate how different smearing functions affect the SNEC constraints on $\Lambda$. We compare the influence of different values of $M$ using both Gaussian and Lorentzian smearing functions, given by Eqs. (\ref{eq:smearGaussian}) and (\ref{eq:smearLorentzian}), respectively. The choice of $M$ allows us to more clearly highlight how the functional shape of the smearing affects the resulting constraints.

\begin{figure}[h]
        \centering
        \includegraphics[width=0.5\linewidth]{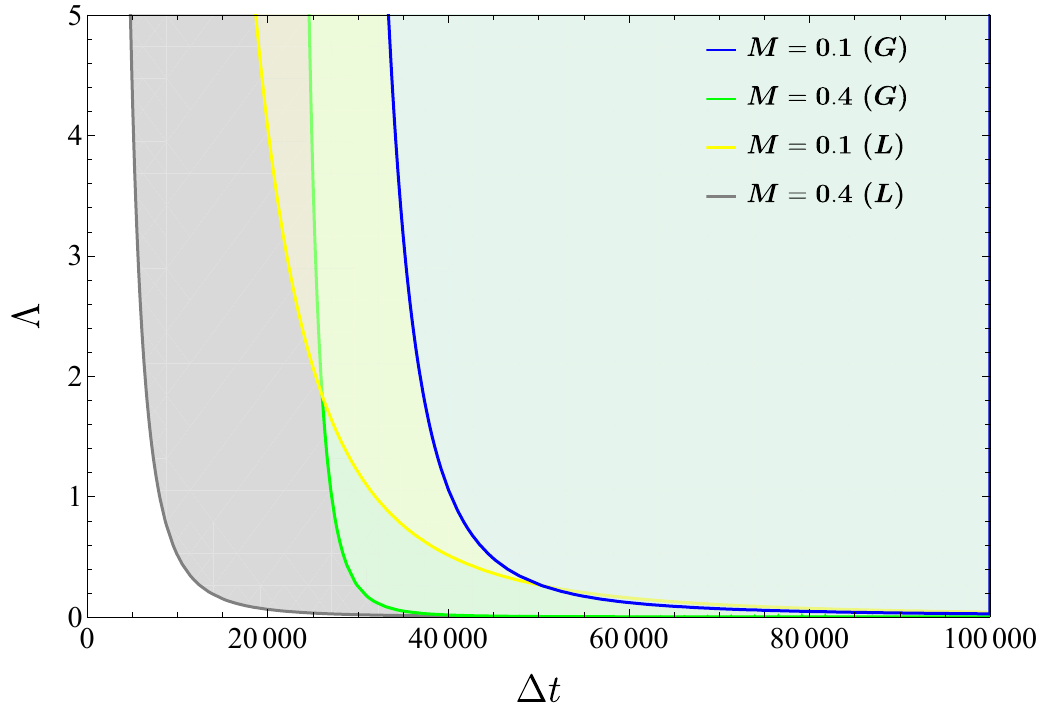}
        \caption{The SNEC constraints with $\bar{t} = -50000$ and $B = 1/(32\pi)$. The labels ``G'' and ``L'' in parentheses indicate the use of Gaussian and Lorentzian smearing functions, respectively. The shaded regions represent the parameter space in which the SNEC is violated.}
        \label{F3}
    \end{figure}

Due to the axis range shown in Fig. \ref{F3}, it is difficult to decisively determine which smearing function yields tighter constraints on $\Lambda$. In principle, if the full parameter space were shown, the constrained regions would cover equal areas since all smearing functions are normalized. Nevertheless, one interesting observation is that the Lorentzian case exhibits a slower growth in the strength of the constraint on $\Lambda$ as $\Delta t$ increases, compared to the Gaussian case.

This behavior can be well understood. For the same smearing width $\sigma$, Lorentzian functions have higher central weight than Gaussian functions; they are more sharply peaked. Gaussian functions, on the other hand, retain significant weight even at distances of one standard deviation from the mean, whereas Lorentzian functions decay more rapidly. When $\Delta t$ is small, the Lorentzian function's weight remains larger near the center, leading to a tighter constraint on $\Lambda$. However, as $\Delta t$ increases, the Lorentzian's weight decreases quickly, and the broader Gaussian distribution eventually contributes more to the integral. This explains why the growth in the strength of the constraint on $\Lambda$ is slower for the Lorentzian case. It is therefore expected that, in the regime of large $\Delta t$, the Gaussian function will yield tighter constraints than the Lorentzian.

The enhancement of the constraint on $\Lambda$ becomes more pronounced near the corner, since the Gaussian smearing function increases more rapidly and therefore produces a sharper turning behavior. In contrast, the Lorentzian smearing function grows more slowly, resulting in a correspondingly smoother corner.

This analysis is meaningful when using the SNEC to constrain physical parameters in comparison with observational data. If NEC-violating effects are sharply localized in time and decay rapidly, choosing a Lorentzian smearing function may better reflect observational sensitivity. In contrast, if the NEC-violating phase has a longer duration and lacks a sharply defined observational window, a Gaussian function provides a more robust and inclusive probe.

In a short summary, the SNEC effectively acts as a window that probes NEC violation. It has a certain threshold, and the tighter the window aligns with regions of stronger NEC violation (i.e., larger $\dot{H}$), the more stringent the constraint on the model parameters becomes. This result is consistent with our physical intuition that the NEC cannot be violated without limit.

\subsection{Case II: $\alpha = 2$}

Again, we can obtain the scale factor through Eq. \eqref{eq:Halpha2} as
\begin{equation}
\label{eq:aalpha2}    a(t)=\mathrm{exp}\left[\frac{1+2\kappa^2\gamma}{80\kappa^2M^2(-t)^4}\right]\simeq1+\frac{1+2\kappa^2\gamma}{80\kappa^2M^2(-t)^4}\,.
\end{equation}
To apply the SNEC conjecture to constrain the Genesis universe in the case $\alpha=2$, we likewise need to introduce an end time $t_{\mathrm{e}}$ for the Genesis phase. From Eq. (\ref{eq:aalpha2}), we obtain the condition for the end of the Genesis phase, leading to the cutoff time:
\begin{equation}
    t_{\mathrm{e}}=-\left(\frac{1+2\kappa^2\gamma}{80\kappa^2M^2}\right)^{\frac{1}{4}}\,,
\end{equation}
We then proceed in the same manner as in Case I: the smearing function is normalized over the interval $(-\infty, t_{\mathrm{e}})$, and the upper integration limit $t_{+}$ is chosen analogously to that in Eq. \eqref{eq:t_+}.

In this part, we also adopt a Gaussian smearing function and numerically evaluate the SNEC conjecture. The result depends on the values of $\bar{t},~\Delta t,~B,~\kappa,~\gamma$ and $M$. We separately examine the impact of $\Delta t$, $\bar{t}$, and $M$ on the parameter spaces $\kappa$ and $\gamma$, and present each case in turn.
When applying the SNEC conjecture, the dependence on $B$ is manifest with the expression in \eqref{eq:SNECdt}. In case I we have also performed a numerical analysis, with the results shown in Fig. \ref{F2}: smaller values of $B$ lead to more stringent constraints on the parameter space.

\begin{figure}[h]
    \centering
    \includegraphics[width=0.5\linewidth]{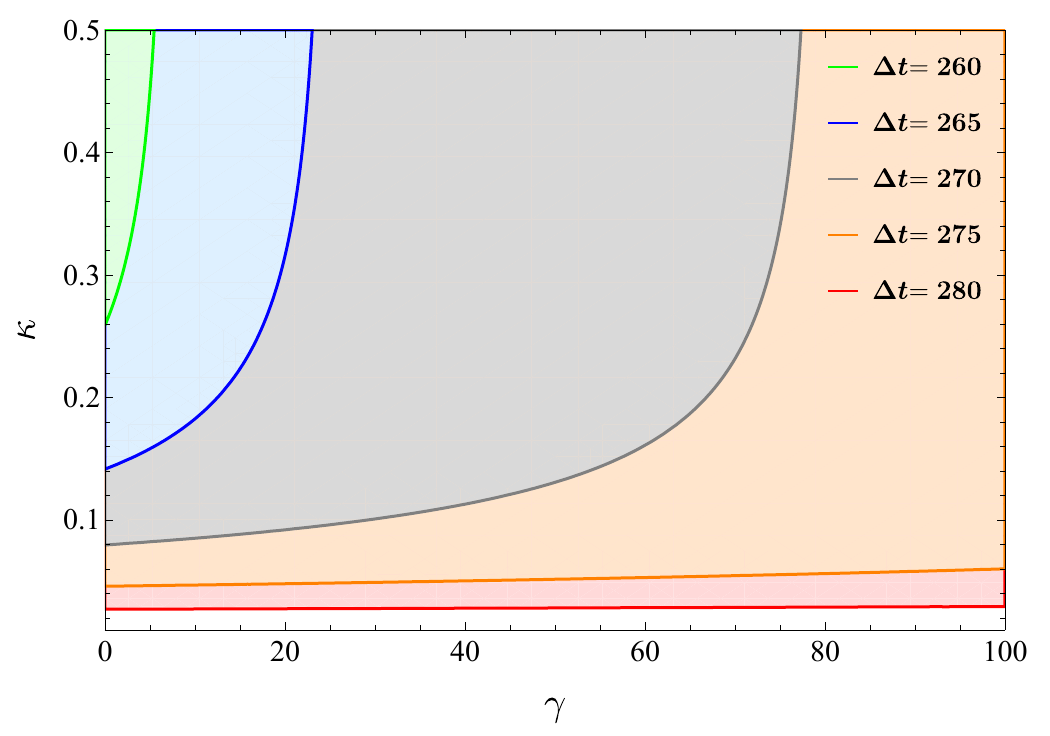}
    \caption{The SNEC constraints with $B = 1/(32\pi)$, $M = 0.9$ and $\bar{t} = -500$ fixed. The shaded regions indicate the parts of parameter space in which the SNEC is violated.}
    \label{F4}
\end{figure}

First, we show the constraints imposed by $\Delta t$ on the parameter space, as illustrated in Fig. \ref{F4}.
A larger shaded region indicates that the SNEC imposes stronger constraints on the parameter space. In both the $\alpha=1$ case and the $\alpha=2$ case, we find that the SNEC constraints on the parameter space become more stringent as $\Delta t$ increases. As mentioned above, $\Delta t$ represents the width of the window in the smearing procedure, and the Genesis universe violates the NEC throughout the whole evolution. Therefore, a larger $\Delta t$ necessarily leads to more stringent constraints.

\begin{figure}[h]
    \centering
    \includegraphics[width=0.5\linewidth]{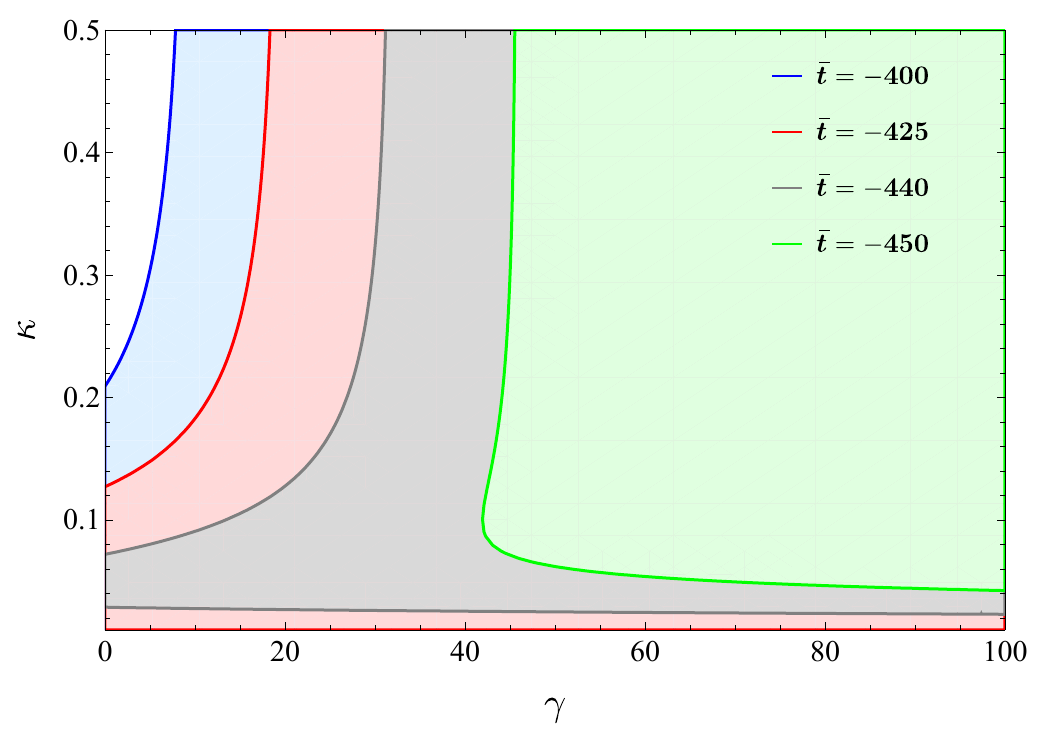}
    \caption{The SNEC constraints with $B = 1/(32\pi)$, $M = 0.9$ and $\Delta t=250$ fixed. The shaded regions indicate the parts of parameter space in which the SNEC is violated.}
    \label{F5}
\end{figure}

Now we move to the constraints on the parameter space with different values of $\bar{t}$. The result is shown in Fig. \ref{F5}.
We see that as $\bar{t}$ approaches $t_{\mathrm{e}}$, the SNEC constraints on the parameter space become more stringent. This is because $\bar{t}$ controls the center of the smearing region. At $\bar{t}$, the smearing function attains its maximal weight, so the magnitude of NEC violation at this point has the largest impact on the resulting SNEC bound. From the expression of $\dot{H}$ in \eqref{eq:Halpha2}, we find that it increases as one approaches $t_{\mathrm{e}}$, implying that the magnitude of NEC violation also increases. Therefore, the closer $\bar{t}$ is to $t_{\mathrm{e}}$, the more stringent the SNEC constraints become.

It is interesting to note that different values of $\Delta t$ and $\bar{t}$ induce different constraints on, and have different impacts on, the parameter space. This is because the parameters $\gamma$ and $\kappa$ affect both the magnitude of NEC violation and the end time $t_{\mathrm{e}}$ in different ways. As for the magnitude of NEC violation, namely the value of $\dot{H}$, the effect of $\gamma$ is such that a larger $\gamma$ leads to a larger $\dot{H}$, i.e. a more pronounced violation of the NEC. In contrast, the effect of $\kappa$ on $\dot{H}$ is the opposite.

However, their impact on the end time $t_{\mathrm{e}}$ is reversed: a larger $\gamma$ makes the Genesis phase end earlier, while a larger $\kappa$ delays the end time $t_{\mathrm{e}}$. Since $\dot{H} \propto 1/(-t)^6$, the NEC violation becomes stronger as one approaches $t_{\mathrm{e}}$. An earlier end time $t_{\mathrm{e}}$ therefore truncates the late-time stage in which the NEC violation is strongest, thereby relaxing the constraints. The contributions of $\gamma$ and $\kappa$ to $\dot{H}$ and to $t_{\mathrm{e}}$ thus act in opposite directions, leading to a competition between these effects, which is precisely what underlies the results shown in Fig. \ref{F4} and Fig. \ref{F5}.

This behavior can be understood from a simple analysis. When studying the impact of $\Delta t$ on the parameter space, we keep the center $\bar t$ fixed. This implies that the weights assigned to the later-time region are relatively small, so their contribution is much less important than the effect of the end time $t_{\mathrm{e}}$ in $\dot{H} \propto 1/(-t)^6$. For the parameter choices considered here we take $\Delta t > 250$ (while cases with $\Delta t \lesssim 250$ are allowed within the plotting range), so that $t_{+}$ is always larger than $t_{\mathrm{e}}$. A later end time $t_{\mathrm{e}}$ therefore leads to stronger SNEC constraints on the parameter space. Consequently, in Fig. \ref{F4}, as we decrease $\Delta t$, the region where the SNEC is violated tends to move towards larger $\kappa$ and smaller $\gamma$, which correspond to a later end time.

When studying the dependence on $\bar t$, the values of $\bar t$ and $\Delta t$ are chosen such that the resulting $t_{+}$ is always larger than $t_{\mathrm{e}}$. This means that the end time $t_{\mathrm{e}}$ remains important for the SNEC bound. At the same time, changing $\bar t$ also shifts the center of the smearing function; since the weight there is large, the SNEC constraint becomes sensitive to the value of $\dot{H}$ at that time. As discussed above, the parameters $\gamma$ and $\kappa$ have competing effects. One eventually finds that, when $\bar t$ is chosen appropriately with respect to $t_{\mathrm{e}}$, the region where the SNEC is violated tends to move towards larger $\kappa$, corresponding to a later $t_{\mathrm{e}}$ but a smaller $\dot{H}$. These tendencies compete with each other. It may be that, for the particular parameter choices adopted here, the effects associated with larger $\gamma$ and larger $\kappa$ dominate, but this is not entirely transparent analytically. The detailed structure of the allowed parameter space must be determined numerically, as shown in Fig. \ref{F5}.

We find that, for the choice of parameters $\gamma = 72$ and $\kappa = 1/12$ adopted in \cite{Zhu:2023lbf}, our numerical analysis reveals a finite region in the ($\Delta t,~\bar{t},~M$) parameter space in which the SNEC is not violated. This, in turn, allows one to construct a stable, non-singular cosmological model consistent with the SNEC conjecture.

\begin{figure}[h]
    \centering
    \includegraphics[width=0.5\linewidth]{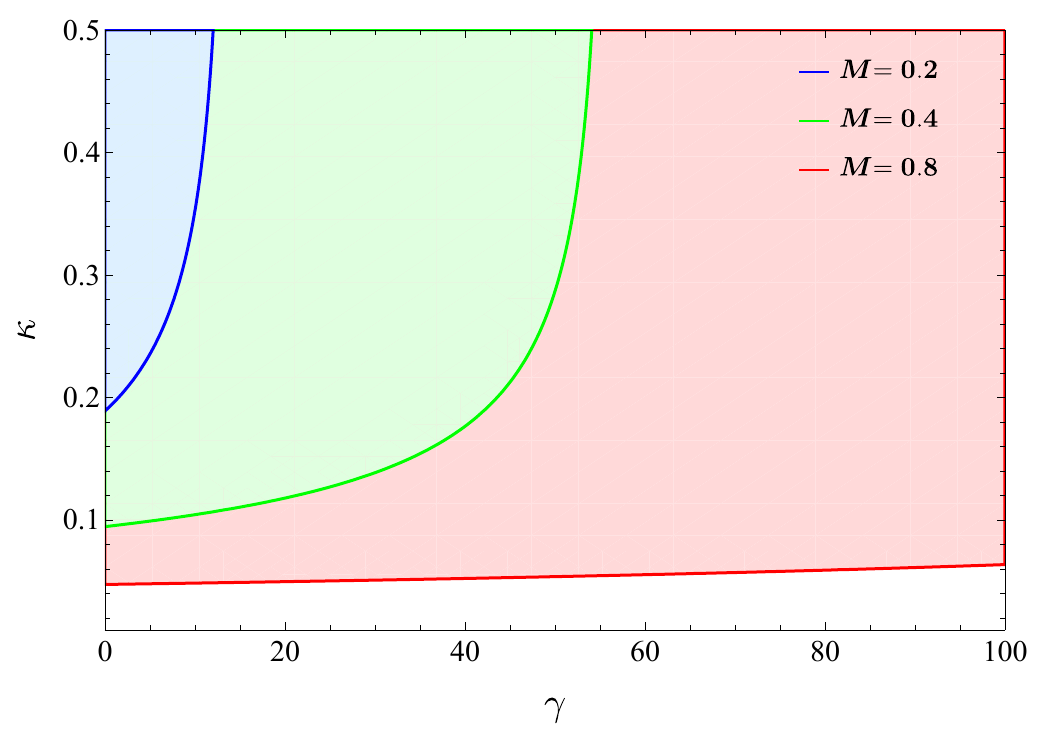}
    \caption{The SNEC constraints with $B = 1/(32\pi)$, $\bar{t} = -450$ and $\Delta t=250$ fixed. The shaded regions indicate the parts of parameter space in which the SNEC is violated.}
    \label{F6}
\end{figure}

Finally, we discuss the constraint with different values of $M$. The result is shown in the Fig. \ref{F6}.
The parameter $M$ also affects the magnitude of NEC violation, i.e., $\dot{H}$, and the end time $t_{\mathrm{e}}$ in opposite ways: a larger $M$ leads to a later $t_{\mathrm{e}}$ but a smaller $\dot{H}$. The analysis of $M$ is relatively straightforward, since in this case we keep $\Delta t$ and $\bar{t}$ fixed. For $\dot{H}$, the NEC violation at $\bar{t}$ is much weaker than that near the end of the Genesis phase. In the vicinity of the end time one has $\dot{H} \propto 1/\left(M^2(-t)^6\right)$, so the dependence of $\dot{H}$ on $M$ is much less significant than its dependence on $t_{\mathrm{e}}$. This explains why, as $M$ decreases, the region where the SNEC is violated tends to move towards comparable values of $\kappa$ but smaller $\gamma$, which correspond to a later end time $t_{\mathrm{e}}$.

\section{Conclusion}
\label{sec:conclusion}

In this work, we have investigated the viability of Genesis cosmology in light of the SNEC conjecture. While the violation of the classical NEC is a fundamental requirement for the Genesis scenario to emerge from a Minkowski state without an initial singularity, the extent of such violation must be physically bounded. The SNEC offers a quantum-motivated, semi-local restriction on NEC violation, serving as a criterion to ensure that the accumulation of negative energy remains within physically reasonable limits.

We applied the SNEC bound to two representative realizations of Genesis cosmology constructed within Generalized Galileon theories, specifically the models corresponding to $\alpha=1$ and $\alpha=2$ as summarized in Ref. \cite{Nishi:2015pta}. We examined the relationship between the model parameters and the characteristics of the smearing window. Our analysis indicates that the constraints are sensitive to both the width of the smearing scale and the specific time interval where the NEC violation is most pronounced.

Ultimately, our results demonstrate that the SNEC imposes nontrivial and significant restrictions on the parameter space of Genesis models. We found that for a stable Genesis phase to exist consistent with the SNEC, the coupling constants and the duration of the NEC-violating phase must fall within specific viable regions. This study highlights that the SNEC conjecture is a powerful and effective tool for constraining nonsingular cosmological scenarios.

\acknowledgments

We thank Yun-Song Piao for valuable discussions.
Dong-Hui Yu and Yong Cai are supported in part by the National Natural Science Foundation of China (Grant Nos. 12575066, 11905224) and the Natural Science Foundation of Henan Province (Grant No. 242300420231). Mian Zhu is supported by the National Natural Science Foundation of China (Grant No. 12503005), and the Fundamental Research Funds for the Central Universities Grant No. YJ202551.
The data that support the findings of this article are openly available \cite{dataYu260228}.

\appendix
 
\section{Consistency of SNEC with perturbative secular growth in de Sitter space}\label{Sec:app1}

The secular growth of $\langle T_{\mu\nu}k^\mu k^\nu\rangle \propto -\frac{4}{3}\frac{\lambda H^4}{(4\pi)^4}\ln[a(t)]$ for a massless, minimally coupled scalar with quartic self-interaction on non-dynamical de Sitter background (as computed in \cite{Kahya:2009sz} with appropriate initial state corrections) is well-founded and deserves careful examination. In the following, we check whether this example satisfies the SNEC.

\subsection{A geometric perspective on the SNEC}

The energy condition is of vital importance in cosmology as it connects the physics of spacetime and the matter content. In the context of primordial universe, we are mainly interested in the property of geometry instead of matter content. For instance, the famous ``no-go'' theorem in Ref. \cite{Borde:1993xh, Borde:2001nh} concerns on the geodesic completeness of the spacetime, i.e., whether the geodesics of a particle terminates at a finite past. The proof is purely geometric, as only the FLRW metric is involved. Thus, their result actually states that an FLRW universe satisfying Null Convergence Condition (NCC) must encounter the initial singularity. Till this step, there is nothing to do with the energy condition of matter content. Now working in general relativity, the Einstein equation relates $G_{\mu \nu}$ and $T_{\mu \nu}$, and the violation of NCC is equivalent to NEC violation, and the statement that ``the free of initial singularity requires NEC violation'' is made in this sense.

The similar logic applies to the study of SNEC. In Ref. \cite{Freivogel:2018gxj}, the foundation paper of SNEC, the authors state that ``The quantity that is bounded by this conjecture is related by the Raychadauri equation to the focusing of null geodesics. As we motivate in more detail below, this conjecture claims that whenever geodesics `de-focus' due to negative null energy, they do so by a small amount that remains in the linear gravity regime." (which can be found in the paragraphs below Eq. (4) of that paper). Namely, the SNEC bound is motivated by the property of null geodesics, firstly written as $R_{kk}^{s} \geq -\#/\tau^2$ indicated in Eq. (6) in that paper, and then translated into $T_{kk}^{s} \geq -\#/\tau^2$ using Einstein equation.

In the setup of Ref. \cite{Kahya:2009sz}, the geometry is assumed to be a non-dynamical de Sitter background, which implies that $H$ is a constant. Consequently, it seems that both the NCC (i.e., $R_{\mu \nu} k^{\mu} k^{\nu}\geq 0$) and the SNEC are satisfied by the scenario studied in Ref. \cite{Kahya:2009sz}, since $R_{\mu \nu} k^{\mu} k^{\nu}=-2\dot{H}(k^0)^2=0$ for the spatially flat Friedmann universe.
\\

\subsection{A matter perspective and the backreaction}

For a massless, minimally coupled scalar with quartic self-interaction on a fixed non-dynamical de Sitter background, after applying the initial state corrections proposed in \cite{Kahya:2009sz}, the expectation value is
\begin{equation}
\langle T_{\mu\nu}k^\mu k^\nu\rangle \propto \frac{\lambda H^4}{(4\pi)^4} \times \left(-\frac{4}{3}\right)\ln a + O(\lambda^2)\,,
\label{eq:Tkk}
\end{equation}
which can be derived from Eqs. (3) and (4) of Ref. \cite{Kahya:2009sz}.
This expression indeed violates the classical NEC since $\langle T_{\mu\nu}k^\mu k^\nu \rangle < 0$ for $a(t) > 1$. As the universe expands, the accumulated negative energy grows logarithmically with the scale factor ($\propto \ln a$) and it appears sufficient to violate the SNEC eventually. However, in a self-consistent theory of semiclassical gravity, one must consider the backreaction of these quantum fluctuations on the spacetime geometry. We demonstrate below that the backreaction becomes non-negligible well before the accumulated negative energy is sufficient to violate the SNEC.

To make this argument quantitative, we compare the timescale (or $e$-folding number) required to violate SNEC ($N_{\text{SNEC}}$) against the timescale at which the background geometry significantly deviates from de Sitter space ($N_{\text{BR}}$).

\subsubsection{The Timescale for SNEC Violation}

First, we derive the condition for SNEC violation. Consider a flat FLRW metric $\d s^2 = -\d t^2 + a^2(t)\d\vec{x}^2$. Let $k^\mu = (k^0, k^i)$ be a null vector tangent to the geodesic $\gamma(\lambda)$ with affine parameter $\lambda$, where $k^0 = a |\vec{k}|$ and $\d t/\d\lambda = k^0$. We have $\langle T_{\mu\nu} k^\mu k^\nu \rangle = (\rho+p)(k^0)^2$. According to Eqs.~(3) and (4) of \cite{Kahya:2009sz}, for a $\lambda \phi^4$ theory, 
\begin{equation}|(\rho+p)_Q| \approx \frac{\lambda H^4}{(4\pi)^4} \left( \frac{4}{3} N \right) \, ,
\end{equation}
where $N = \ln a$ is the $e$-folding number, the subscript ``Q'' denotes contributions from quantum corrections.

The SNEC requires $\int \d\lambda f^2(\lambda) \langle T_{\mu\nu} k^\mu k^\nu \rangle \geq - B / (G \sigma^2)$, where $\sigma$ is the smearing scale in affine units. Since $\ln a$ varies slowly over a Hubble time ($\Delta N \sim 1 \ll N$ at late times), we can approximate the integral by the local value. To obtain the most physically relevant constraint, the physical smearing time $\Delta t$ is chosen to be of the order of the curvature scale, $\Delta t \sim H^{-1}$ (see e.g. Ref. \cite{Moghtaderi:2025cns}). The corresponding affine smearing scale is $\sigma \approx \Delta \lambda = \Delta t / k^0 \approx (k^0 H)^{-1}$. Substituting these into the SNEC inequality, we find \begin{equation}
(\rho+p)(k^0)^2 \gtrsim - \frac{B}{G [ (k^0 H)^{-1} ]^2} = - \frac{B (k^0)^2 H^2}{G} \, .
\end{equation}
We observe that the factor $(k^0)^2$ cancels out on both sides, yielding a bound: \begin{equation}
|(\rho+p)| \lesssim 8\pi B M_{\mathrm{P}}^2 H^2\,.
\end{equation} 
Note that we have used $8\pi G = M_{\mathrm{P}}^{-2}$. Equating the quantum source term to this bound, we find the timescale $N_{\text{SNEC}}$:
\begin{equation}\frac{\lambda H^4}{(4\pi)^4} \frac{4}{3} N_{\text{SNEC}} \approx 8\pi B M_{\mathrm{P}}^2 H^2 \quad \Longrightarrow \quad N_{\text{SNEC}} \approx \frac{6 \pi B (4\pi)^4}{\lambda} \left( \frac{M_{\mathrm{P}}}{H} \right)^2 \, .
\end{equation}
Therefore, the SNEC will be violated when the universe expands to the scale where $\ln a = N_{\text{SNEC}}$.

\subsubsection{Comparison with Backreaction Timescales}

We now evaluate when the fixed background approximation breaks down. We consider two distinct criteria for the onset of significant backreaction.

{\it Criterion I: Geometric Deviation (Slow-roll).} We can define backreaction based on the induced time-evolution of the Hubble parameter $H$. Using $2M_{\mathrm{P}}^2|\dot{H}| = |(\rho+p)|$, we define $N_{\text{BR1}}$ at the time when the parameter $\mathcal{E} \equiv |\dot{H}/H^2|$ reaches a threshold $\mathcal{E}_0$, i.e.,
\begin{equation}
\frac{\lambda H^4}{(4\pi)^4} \frac{4}{3} N_{\text{BR1}} = 2 \mathcal{E}_0 M_{\mathrm{P}}^2 H^2 \quad \Longrightarrow \quad N_{\text{BR1}} \approx \frac{3 \mathcal{E}_0 (4\pi)^4}{2 \lambda} \left( \frac{M_{\mathrm{P}}}{H} \right)^2 \, .
\end{equation}
The ratio of timescales is
\begin{equation}
\frac{N_{\text{SNEC}}}{N_{\text{BR1}}} = \frac{4\pi B}{\mathcal{E}_0} \, .
\end{equation}
For typical values of those parameters (e.g., $B \approx 1/32\pi$ and a conservative $\mathcal{E}_0 \sim 10^{-1}$), this ratio is greater than unity ($N_{\text{SNEC}} \gtrsim N_{\text{BR1}}$). This indicates that when the SNEC is violated, the evolution of $H$ over time has become sufficiently significant such that the fixed background approximation is nearly broken down.

{\it Criterion II: Energy Density Dominance.} The leading order contribution to the energy density grows quadratically with $N$. From Eq.~(3) of \cite{Kahya:2009sz}, $\rho_Q \approx \frac{\lambda H^4}{(4\pi)^4} 2 N^2$. We define the backreaction time $N_{\text{BR2}}$ as the moment when the quantum energy density reaches a fraction $\varepsilon$ of the background energy density $3 M_{\mathrm{P}}^2 H^2$, i.e.,
\begin{equation}
\frac{\lambda H^4}{(4\pi)^4} 2 N_{\text{BR2}}^2 = \varepsilon \cdot 3 M_{\mathrm{P}}^2 H^2 \quad \Longrightarrow \quad N_{\text{BR2}} \approx \sqrt{\frac{3 \varepsilon (4\pi)^4}{2 \lambda}} \left( \frac{M_{\mathrm{P}}}{H} \right) \, .
\end{equation}
Comparing the two timescales, we find
\begin{equation}
\frac{N_{\text{SNEC}}}{N_{\text{BR2}}} \approx \pi B \sqrt{\frac{72}{\varepsilon \lambda}} \left( \frac{M_{\mathrm{P}}}{H} \right) \gg 1 \,.
\end{equation}
Since $M_{\mathrm{P}}/H$ is typically very large (e.g., $10^5$), $B \approx 1/32\pi$, and $\rho_Q$ grows as $N^2$ while $|(\rho+p)_Q|$ grows only as $N$, the backreaction from the energy density will destroy the de Sitter background long before the SNEC bound is challenged.

Therefore, the SNEC is satisfied by the example given by Ref. \cite{Kahya:2009sz} from both a geometric perspective and a matter perspective. Intriguingly, the backreaction becomes significant before the SNEC is violated. As has been pointed out in the footnote on page 3 of \cite{Kahya:2009sz}, ``these secular corrections eventually become nonperturbatively strong''. The result given by Eq. (\ref{eq:Tkk}) relies on the assumption of a background that remains fixed despite the growth in the stress-energy tensor. In a physically realistic universe, the geometry would evolve (breaking the de Sitter symmetry) well before the secular growth of the NEC-violating term could break the SNEC bound. Therefore, the SNEC is a consistent constraint for dynamical cosmological models.

Additionally, it is also interesting to consider the phenomenon of ``self-reproduction'' during inflation \cite{Linde:1986fd}. While quantum fluctuations can indeed drive the scalar field up its potential in certain regions (increasing local $H$), they simultaneously drive the field down in other regions (decreasing $H$). The SNEC is a constraint on the semi-classical expectation value $\langle T_{\mu\nu} \rangle$ along a geodesic. Those localized fluctuations do not imply that the integrated expectation value $\int \langle T_{\mu\nu} \rangle k^\mu k^\nu f(\lambda) d\lambda$ along a geodesic becomes negatively divergent. The ensemble average of the energy-momentum tensor remains controlled by the competition between classical drift and quantum diffusion. Thus, a local, stochastic increase in $H$ in a ``self-reproducing'' patch does not automatically imply that the semi-classical integral condition is violated for the entire spacetime or for the Genesis scenario we are investigating.

\bibliography{necv}
\bibliographystyle{utphys}

\end{document}